\documentclass[aps,
showpacs,
preprint,
]{revtex4}
\usepackage{epsf}
\begin{document}
\title{Substrate Adhesion of a Nongrafted Flexible Polymer in a Cavity}
\author{Michael Bachmann}
\email[E-mail: ]{Michael.Bachmann@itp.uni-leipzig.de}
\author{Wolfhard Janke}
\email[E-mail: ]{Wolfhard.Janke@itp.uni-leipzig.de}
\homepage[\\ Homepage: ]{http://www.physik.uni-leipzig.de/CQT.html}
\affiliation{Institut f\"ur Theoretische Physik, Universit\"at Leipzig,
Augustusplatz 10/11, D-04109 Leipzig, Germany}
\begin{abstract}
In a contact density chain-growth study we investigate the solubility-temperature pseudo-phase 
diagram of a lattice polymer
in a cavity with an attractive surface. In addition to the main phases of adsorbed and
desorbed conformations we find numerous subphases of collapsed and expanded structures.
\end{abstract}
\maketitle
\section{Introduction}
\label{secintro}
The requirement of higher integration scales in electronic circuits, the onset of nanosensory 
applications in biomedicine, but also the fascinating capabilities of modern experimental setup
with its enormous potential in polymer and surface research recently led to an increasing interest
at the hybrid interface of organic and inorganic matter~\cite{brown1,whaley1,goede1,willett1,gray1}.  
This also includes numerous detailed studies, e.g., of polymer film wetting 
phenomena~\cite{wetting1,wetting2},
pattern recognition~\cite{nakata1,bogner1}, protein--ligand binding and 
docking~\cite{ligand1,ligand2,irbaeck1}, charged adsorbed polymers~\cite{cheng1} 
as well as deposition and growth of polymers at surfaces~\cite{foo1}. 

In most theoretical and computational studies the polymer is anchored at the substrate with one 
of its ends which reduces the entropic freedom of the polymer. These surface-grafted 
polymers~\cite{grassberger1,vrbova1,singh1,causo1,prellberg1,huang1}
are, e.g., of particular interest in studies of shape transformations~\cite{lipowsky1}, e.g., 
as reaction to external fields~\cite{frey1,celestini1,prellberg2}. However, in many recent 
experiments of organic--inorganic interfaces the setup is different~\cite{whaley1,goede1} and is 
more adequately described by a polymer moving in a cavity with one adsorbing surface~\cite{bj1,bj2}. 
The main difference of such nongrafted polymers considered in this work 
is of entropic kind: In the desorbed phase the polymer can move freely within 
the cavity, and the polymer can fold into conformations, where the ends have no contact with the 
surface. 

This paper is organized as follows. In Sect.~\ref{secmod} we describe the details of a minimalistic
model for the hybrid system. The main result, the solubility-temperature 
pseudo-phase diagram, is presented and discussed in Sect.~\ref{sechyb}. The interpretation
is consolidated by exemplified studies of fluctuations and correlations of relevant thermodynamic quantities 
such as numbers of contacts between monomers and monomer-substrate contacts as well as 
the gyration tensor, in the different phases. The contact numbers turn out to be adequate system parameters
for the description of the macrostate of the system, and therefore the free energy in dependence of these
contact numbers is subject of a detailed study in Sect.~\ref{secfree}. This quantity is also useful for 
classifying the conformational transitions between the phases which are also discussed there. Eventually, 
we conclude in Sect.~\ref{secsum} with a summary of the main results.
\section{Minimalistic model for polymer adsorption}
\label{secmod}
\begin{figure}
\centerline{\epsfxsize=7cm \epsfbox{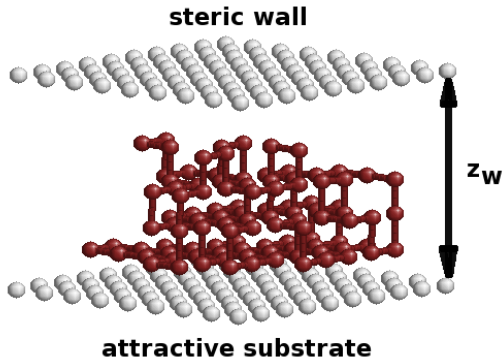}}
\caption{\label{figmod} Cavity model used in this work. The lower of the two parallel
surfaces is attractive to the polymer, the upper is steric only. The distance between the
surfaces is $z_w$ lattice units.} 
\end{figure}

We employ a minimalistic simple-cubic (sc) lattice model~\cite{bj1} which allows a systematic 
analysis of the
conformational phases experienced by a nongrafted polymer in a cavity with one adhesive 
surface. An example for the cavity model is shown in Fig.~\ref{figmod}. The polymer can move between the
two infinitely extended parallel planar walls, separated by a distance $z_w$ expressed in lattice units.  
The substrate is short-range attractive to the monomers of the polymer chain, while the influence
of the other wall is purely steric. 

Denoting the number of nearest-neighbor, but nonadjacent monomer-monomer contacts by $n_m$ and
the number of nearest-neighbor monomer-substrate contacts by $n_s$, the energy of the hybrid
system can be expressed in the simplest model as
\begin{equation}
\label{eqenergy}
E_s(n_s,n_m)=-\varepsilon_s n_s-\varepsilon_m n_m,
\end{equation}
where $\varepsilon_s$ and $\varepsilon_m$ are the respective contact energy scales, which are left 
open in the following. For simplicity, we perform a simple rescaling and set 
$\varepsilon_s=\varepsilon_0$ and $\varepsilon_m=\varepsilon_0 s$. Here we have introduced the 
overall energy scale $\varepsilon_0$ and the dimensionless
reciprocal solubility $s$ that controls the quality of the implicit solvent surrounding the polymer
(the larger the $s$, the worse the solvent). 
Since contacts with the substrate usually
entail a reduction of monomer-monomer contacts, there are two competing forces (rated against
each other by the energy scales) affecting the formation of intrinsic and surface contacts.      
In this paper we mainly focus on the conformational transitions the polymer experiences under
different environmental conditions. Concretely, we are interested in the dependence of energetic and 
structural quantities on temperature $T$ and reciprocal solubility $s$ in equilibrium.
The probability (per unit area) for a conformation with $n_s$ surface and $n_m$ monomer-monomer contacts
at temperature $T$ and reciprocal solubility $s$ is given by
\begin{equation}
\label{eqprob}
p_{T,s}(n_s,n_m)=\frac{1}{Z}g_{n_s n_m}e^{\varepsilon_0(n_s+s n_m)/k_BT},
\end{equation}
where
$g_{n_s n_m} = \delta_{n_s 0}\,g^{\rm u}_{n_m}+(1-\delta_{n_s 0})g^{\rm b}_{n_s n_m}$
is the contact density and $Z$ the partition sum. In this decomposition, $g^{\rm u}_{n_m}$ stands for the density
of unbound conformations, whereas
$g^{\rm b}_{n_s n_m}$ is the density of surface and intrinsic contacts of all
conformations bound to the substrate. Obviously, the number of the conformations without contact to the
attractive substrate, $g^{\rm u}_{n_m}$, depends on the distance $z_w$ between the cavity walls.
For a sufficiently large distance
$z_w$ from the substrate the influence of the neutral surface on the unbound polymer is small.
For $z_w\to\infty$, however, $g^{\rm u}_{n_m}$ formally diverges. Therefore, the non-adhesive, 
impenetrable steric wall is necessary for regularization.

We studied polymers with up to 200 monomers by applying the contact-density chain-growth
algorithm which is an improved variant of the recently developed multicanonical
chain-growth sampling method~\cite{bj3,prellberg3}. All these methods set up on
a variant of the pruned-enriched variant~\cite{grassberger2} of Rosenbluth 
sampling~\cite{rosenbluth1}. The main advantage
of the improved method is that it directly samples the contact density $g_{n_s n_m}$,
which is very useful for problems, where the model provides different
energy scales. This generalizes the ordinary multicanonical version~\cite{bj3} which samples
the density of states, i.e., the number of states for given energy. Here we can set
the two independent energy scales $\varepsilon_m$ and $\varepsilon_s$ or their
ratio $s$, respectively, {\it after} the simulation. This allows to introduce the
reciprocal solubility $s$ as a second environmental parameter in addition to the temperature $T$.
 
The partition sum of the system as a function of these two parameters is simply 
$Z=\sum_{n_s,n_m} g_{n_s n_m}\exp\{\varepsilon_0(n_s+s n_m)/k_BT\}$ and 
the statistical average of any function $O(n_s,n_m)$ is given by the formula
\begin{equation}
\label{eqav}
\langle O\rangle(T,s)=\sum_{n_s,n_m}O(n_s,n_m)p_{T,s}(n_s,n_m), 
\end{equation}
which is very convenient since it only requires to estimate the contact density 
$g_{n_s n_m}$ in the simulation. Denoting contact correlation matrix elements as 
$M_{xy}(T,s)=\langle xy\rangle_c=\langle xy\rangle-\langle x\rangle\langle y\rangle$
with $x,y=n_s,n_m$, the specific heat can be written as
\begin{equation}
\label{eqspec}
C_V(T,s) = k_B\left(\frac{\varepsilon_0}{k_BT}\right)^2(1,s)\, M(T,s)
\left(\begin{array}{c}1\\ s\end{array}\right).
\end{equation} 
All quantities depending only on the contact numbers $n_m$ and $n_s$ can therefore
simply be calculated from the estimate of the contact density $g_{n_s n_m}$ provided
by our simulation method. 

Although the two contact parameters are sufficient
to describe the macrostate of the system and their fluctuations characterize the
main pseudo-phase transition lines, it is often useful to introduce also nonenergetic
quantities such as the end-to-end distance and the gyration tensor for gaining more
detailed structural information of the polymer. For our specific problem at hand it is
particularly useful to study the structural anisotropy of the adsorbed polymer
in the different phases. To this end, we define the general gyration tensor for
a polymer chain of $N$ beads with the components
\begin{equation}
\label{eqgyr}
R^2_{ij} = \frac{1}{N}\sum\limits_{n=1}^N \left(x_i^{(n)}-\overline{x_i}\right)
\left(x_j^{(n)}-\overline{x_j}\right),
\end{equation} 
where $x^{(n)}_i$, $i=1,2,3$, is the $i$th Cartesian coordinate of the $n$th monomer and
$\overline{x_i}=\sum_{n=1}^N x_i^{(n)}/N$ is the center of mass with respect to the $i$th 
coordinate. Anisotropy in the polymer fluctuations is connected with the system's
geometry and therefore it will be sufficient to study the components of the gyration
tensor parallel ($x$, $y$ components) and perpendicular (in $z$ direction) to the planar walls,
\begin{equation}
\label{eqgyrpara}
R^2_{\parallel} = \frac{1}{N}\sum\limits_{n=1}^N \left[\left(x^{(n)}-\overline{x}\right)^2
+\left(y^{(n)}-\overline{y}\right)^2\right]
\end{equation}
and
\begin{equation}
\label{eqgyrperp}
R^2_{\perp} = \frac{1}{N}\sum\limits_{n=1}^N \left(z^{(n)}-\overline{z}\right)^2.
\end{equation}
The gyration radius is then simply the trace of the gyration tensor, 
$R^2_{\rm gyr}={\rm Tr}\, R^2 = \sum_{i=1}^3 R_{ii}^2=R^2_{\parallel}+R^2_{\perp}$. 
The calculation of statistical averages for quantities $R$ that are not necessarily 
functions of
the contact numbers $n_s$ and $n_m$ cannot be performed via Eq.~(\ref{eqav}). In
this case only the more general relation 
$\langle R\rangle=\sum_{\bf X}R({\bf X})\exp\{-E_s({\bf X})/k_BT\}/Z$ holds,
where the sum runs over all polymer conformations ${\bf X}$.
Introducing the accumulated density $R_{\rm acc}(n_s',n_m')=\sum_{\bf X} R({\bf X})
\delta_{n_s({\bf X})\, n_s'}\delta_{n_m({\bf X})\, n_m'}/Z$, where $\delta_{ij}$ is 
the Kronecker symbol, the expectation value can be expressed, however, in a form similar to 
Eq.~(\ref{eqav}):
\begin{equation}
\label{eqavb}
\langle R\rangle=\sum_{n_s,n_m}R_{\rm acc}(n_s,n_m)e^{-E_s(n_s,n_m)/k_BT}.
\end{equation}
The quantity $R_{\rm acc}(n_s,n_m)$ can easily be measured in simulations with 
the contact density chain-growth algorithm. 
In the following we use natural units, i.e., we set $k_B=\varepsilon_0\equiv 1$.
\section{Thermodynamic behavior of the hybrid system}
\label{sechyb}
For our exemplified study of the hybrid system in equilibrium we chose
a polymer with 179 monomers. Since this is a prime number, the polymer is
unable to form perfect cuboid conformations on the sc lattice, as it is, e.g., the case for 
a 100-mer.~\cite{bj1} There we found two low-temperature
subphases dominated by the same $4\times 5\times 5$ cuboid. In one subphase 
it had 20 surface contacts, while in the other the cuboid was simply 
rotated, entailing 25 surface contacts. This is a typical example, where
the exact number of monomers in the linear chain is directly connected with the 
occurrence of such specific pseudo-phases which are not, of course, phases in the
traditional view. Nonetheless, the enormous progress in high-resolution
experimental structure analyses
and in the technological equipment for precise polymer deposition, as well as 
the natural finite length of classes of polymers (e.g., peptides and proteins),
explain the growing interest in pseudo-phases and the conformational transitions
between them.  
Here we mainly focus on the expected thermodynamic phase transitions~\cite{vrbova1,singh1}  
and low-temperature higher-order layering pseudo-phase transitions~\cite{prellberg1}.
The following results were obtained from contact-density chain-growth simulations of the 
179-mer in a cavity with $z_w=200$ (see Fig.~\ref{figmod}), choosing uniformly distributed
starting points at random. In eight independent runs 
$1.6\times 10^9$ polymer conformations were generated in total. The resulting 
contact density $g_{n_s n_m}$ and accumulated densities like $R_{\rm acc}(n_s,n_m)$ are 
independent of external parameters such as temperature $T$ and reciprocal solubility $s$. 
Concrete values of statistical quantities for specific parameter settings are obtained 
by simple reweighting as in Eqs.~(\ref{eqprob}) and~(\ref{eqavb}).
\begin{figure}
\centerline{\epsfxsize=8.8cm \epsfbox{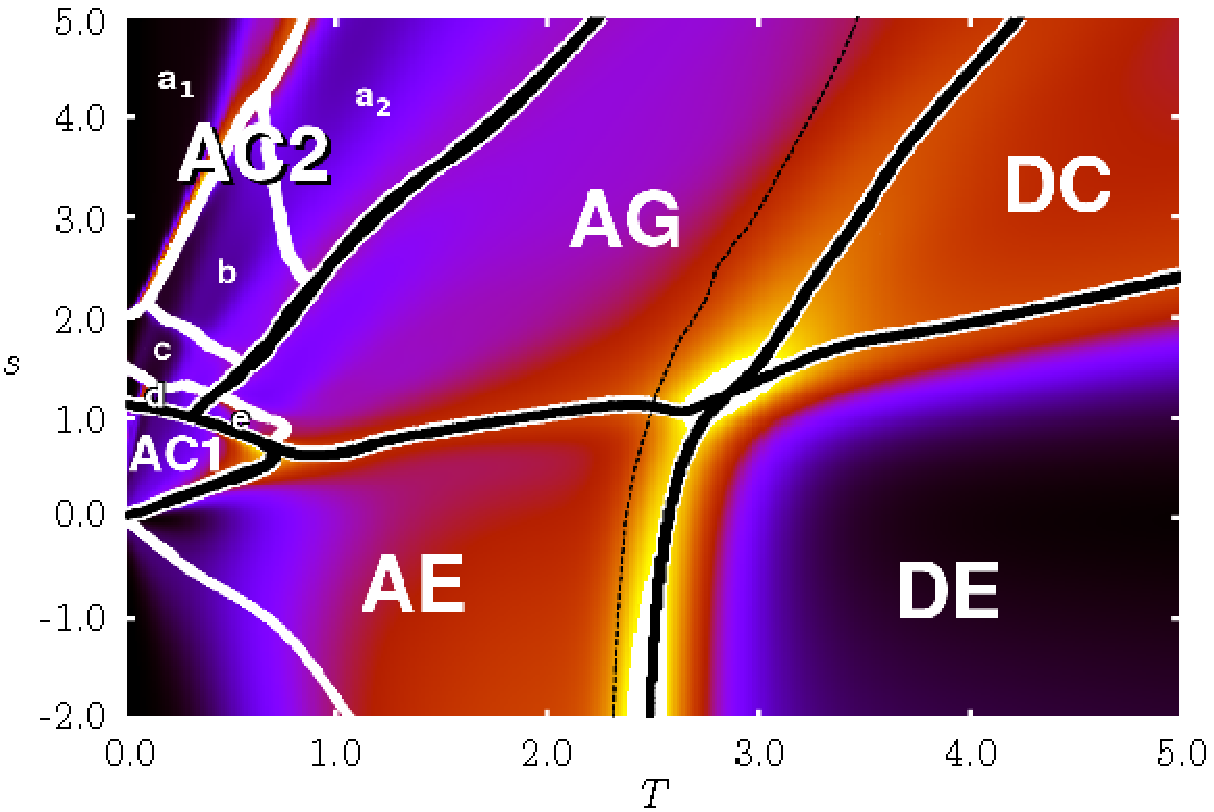}}
\caption{\label{figpd} (Color online) Solubility-temperature pseudo-phase diagram 
of a 179-mer. The color codes the specific heat as a function of reciprocal solubility $s$ and
temperature $T$ -- the brighter the larger its value. Drawn lines emphasize the 
ridges of the profile and indicate transitions between the different conformational
phases. Black lines mark expected thermodynamical phase transitions, while white
lines belong to pseudo-transitions specific to finite-length polymers. Along the dashed 
black line coexisting desorbed and adsorbed conformations are equally probable.} 
\end{figure}
\subsection{Solubility-temperature pseudo-phase diagram}
\label{ssectpd}
Discontinuities or divergences of energetic and nonenergetic fluctuations as functions 
of external parameters reveal typically dramatic cooperative transitions in the collective, 
macroscopic behavior of the system's microscopic degrees of freedom in the thermodynamic
limit. These transitions separate then thermodynamically stable phases and the transitions
can uniquely be identified by certain values of the external parameters, e.g., the transition
temperature. Usually, all fluctuations collapse at the same parameter sets. But,
this ``traditional view'' is only true in the thermodynamic limit. 
Finite-size systems usually exhibit a zoo of crossover- or pseudo-transitions, most of
which disappearing in the thermodynamic limit. In special cases, e.g., proteins, where the
specific amino acid sequence is of finite length, no phase transitions in the strict sense
happen at all.
Still, peaks in curves of fluctuating quantities {\em can be} signatures for ``cooperative activity'',
but this is not necessarily indicated by all fluctuations considered, and if, then typically
at different parameter values.~\cite{bj3} Nonetheless, in protein science,
pseudo-transitions such as conformational transitions are important in the understanding
of secondary structure formation and the tertiary hydrophobic-core collapse. For
polymers, mainly the $\Theta$ collapse transition, which is probably of second order, is
of particular interest.~\cite{deGennes1,grassberger2} This is a real thermodynamic 
phase transition. Nonetheless, at least for finite systems, an additional first-order-like
glassy or crystallization transition at lower temperatures is also conjectured for 
polymers.~\cite{rampf1}  

All these peculiarities of finite polymer systems are also relevant for the adsorption problem
we consider here. In Fig.~\ref{figpd} we have plotted the projection of the specific 
heat profile onto the solubility-temperature plane as obtained from our simulation of the 
179-mer in a cavity with $z_w=200$. The color code reflects the value of the specific heat and
the brighter the shading, the larger the value of $C_V$. Black and white lines emphasize
the ridges of the profile. Since we consider the specific heat as appropriate to identify 
pseudo-phases, these ridges mark the pseudo-phase boundaries. As expected, the pseudo-phase
diagram is divided into two main parts, the phases of adsorption and desorption. The two
desorbed pseudo-phases DC (desorbed-compact conformations) and DE (desorbed-expanded
structures) are separated by the collapse transition line which corresponds to the
$\Theta$ transition of the infinite-length polymer which is allowed to extend into the
three spatial dimensions~\cite{remA}. 
The region of the adsorbed pseudo-phases is much more complex,
and little is known about its details, since it is relevant at lower temperatures, where 
conventional Monte Carlo methods with pivot-like updates usually tend to fail. The presence of 
general phases of adsorbed-expanded (AE)~\cite{rem1} and adsorbed-compact (AC1, AC2) conformations was
postulated in adsorption studies of grafted polymers and the existence of an additional 
phase of surface-attached globules (AG)~\cite{rem1} was assumed~\cite{vrbova1,singh1,prellberg1}. 
In a recent study~\cite{prellberg1}, it was argued that the layering transition between AC1 and AC2
is a thermodynamic phase transition. Although the polymer in our study is still relatively
small, we can clearly identify pseudo-phases in Fig.~\ref{figpd} which can be assigned these 
labels, too. Those regions are separated by the black lines indicating the transitions between
them. We expect that these are transitions in the thermodynamic meaning; only the precise
location of the transition lines will still change with increasing length of the polymer. 
Thus, this picture confirms the previously
assumed phases and it provides evidence that the AG phase is indeed there. 
Furthermore, we have also highlighted by white lines transitions between pseudo-phases which will 
probably not survive in the thermodynamic limit. This concerns, e.g., the higher-order
layering transitions among the compact pseudo-phases AC2a$_{1,2}$-d. In the following 
sections we will analyse the properties of the pseudo-phases in more detail. 
\vspace*{-4mm}
\subsection{Contact-number fluctuations}
\label{ssecfl}
The contact numbers $n_s$ and $n_m$ can be considered as system parameters appropriately describing
the state of the system and are therefore useful to identify the pseudo-phases. Peaks and dips in the 
external-parameter dependence of self-correlations $\langle n_s^2\rangle_c$, $\langle n_m^2\rangle_c$ 
and cross-correlations $\langle n_s n_m\rangle_c$ indicate activity in the contact-number fluctuations
and, analysing the expectation values $\langle n_s\rangle$ and $\langle n_m\rangle$ in these active 
regions of the external parameters $T$ and $s$, allow for an interpretation of the 
respective conformational transitions between the pseudo-phases.

\begin{figure}
\centerline{\epsfxsize=8.8cm \epsfbox{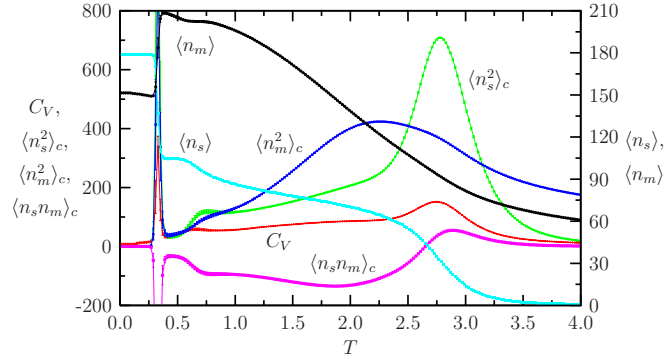}}
\caption{\label{figc179} (Color online) Expectation values, self- and cross-correlations
of the contact numbers $n_s$ and $n_m$ as functions of the temperature $T$ in comparison
with the specific heat for a 179-mer in solvent with $s=1$.} 
\end{figure}
In Fig.~\ref{figc179}, we have plotted for the 179-mer these quantities and, for comparison, the specific
heat as functions of the temperature $T$ at a fixed solvent parameter $s=1$. This example is quite 
illustrative as the system experiences several conformational transitions when increasing the temperature 
starting from $T=0$ (see Fig.~\ref{figpd}). At temperatures very close to $T=0$ (pseudo-phase AC1) 
all 179 monomers have contact
to the substrate and 153 monomer-monomer contacts are formed. This is the most compact contact set
being possible for {\em topologically two-dimensional}, film-like conformations. It should be noted, however, 
that approximately $2\times 10^{18}$ conformations (self-avoiding walks) belong to this contact 
set.~\cite{rem2} This high degeneracy is an artefact of the minimalistic lattice polymer model used.   
It is remarkable that the conformations with the highest number of total contacts $n=n_s+n_m$ are
film-like compact ($n=332$). All other conformations we found possess less contacts, even the most compact 
contact set that dominates the five-layer pseudo-phase AC2a$_1$ ($n_s=36$, $n_m=263$, i.e., $n=299$). The reason is
that for low temperatures, those macrostates are formed which are energetically favored. Entropy is not yet 
relevant -- for the $s=1$ example $\langle n_s\rangle$ drops to 149 only up to $T\approx 0.3$. 
Increasing the temperature further, the situation dramatically changes, 
as can be seen in Fig.~\ref{figc179}. In a highly cooperative process, the average number of intrinsic 
contacts $\langle n_m\rangle$ significantly increases (to $\approx 208$) at the expense of surface 
contacts ($\langle n_s\rangle$ drops to approximately 104). Consequently, the strong fluctuations 
$\langle n_{s,m}^2\rangle_c$ signalize a conformational transition, and the anticorrelation indicated
by $\langle n_s n_m\rangle_c$ confirms that surface contacts turn into intrinsic contacts, which indirectly
leads to the conclusion that the film-like structure is given up in favor of layered, spatially 
three-dimensional conformations. The system has entered subphase AGe which is the part of the phase AG, where
two-layer conformations dominate. The subphase transition near $T\approx 0.7$ from the two-layer (AGe) 
to the bulky regime of AG is due to the ongoing, rather unstructured expansion of the polymer into the 
$z$ direction by forming so-called surface-attached globules~\cite{singh1}. This is accompanied by a further
reduction of surface contacts, while the number of intrinsic contacts changes weakly. Approaching 
$T\approx 2.0$, the situation is just vice versa. Intrinsic contacts dissolve and the system experiences 
a conformational phase transition from globular conformations in AG to random strands in AE. 
Crossing this transition line, the system enters the good-solvent regime. Eventually, close to 
$T\approx 2.8$, the polymer unbinds off the substrate. A clear signal is observed in the fluctuations 
of $n_s$, i.e., the number of average surface contacts rapidly decreases. The expanded polymer is ``free'' 
and the influence of {\em both} walls is effectively steric. This phase (AE) is closely related to the typical 
random-coil phase of entirely free and dissolved polymers in good solvent. 

This example shows that a study of the contact number fluctuations is indeed sufficient to qualitatively
identify and describe the conformational transitions between the pseudo-phases of the hybrid system. 
For this reason, $n_s$ and $n_m$ are adequate system parameters playing a similar role as order 
parameters in thermodynamic phase transitions.
\subsection{Anisotropic behavior of gyration tensor components}
\label{ssecgyr}
\begin{figure}
\centerline{\epsfxsize=8.8cm \epsfbox{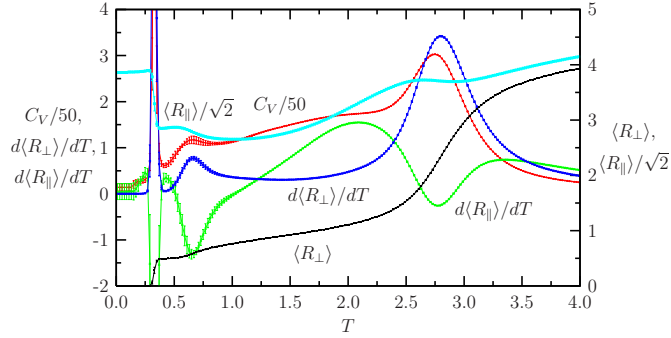}}
\caption{\label{figg179} (Color online) 
Anisotropic behavior of gyration tensor components parallel and perpendicular
to the substrate and their fluctuations as functions of the temperature $T$ 
for a 179-mer at $s=1$. For comparison, we have also plotted the associated specific-heat 
curve.} 
\end{figure}

One of the most interesting structural quantities in studies of polymer phase transitions is the 
gyration tensor~(\ref{eqgyr}). For our hybrid system we expect that the respective components 
parallel~(\ref{eqgyrpara}) and perpendicular~(\ref{eqgyrperp}) to the substrate 
will behave differently when the polymer passes pseudo-transition lines. In order to prove this
anisotropy explicitly, we have plotted in Fig.~\ref{figg179} the expectation values, 
$\langle R_{\parallel,\perp}\rangle$, and the fluctuations of these two components,
$d\langle R_{\parallel,\perp}\rangle/dT$, again for the polymer in solvent with $s=1$. 
For interpreting the peaks of the
fluctuations, we have also included once more the specific heat curve for comparison. 
The immediate observation is that the temperatures, where one or both gyration tensor components exhibit
peaks, almost perfectly coincide with those of the specific heat. This is a strong confirmation for
the phase diagram in Fig.~\ref{figpd} which is based on the specific heat. Obviously, even for the rather
short polymer with 179 monomers, we encounter the onset of fluctuation collapse near the (pseudo-)phase
transitions. This is very promising for future quantitative finite-size scaling analyses.

At very low temperatures, i.e., in pseudo-phase AC1, we have argued in the previous section that 
the polymer-conformation is the most compact single-layer film. This is confirmed by the behavior 
of $\langle R_\parallel\rangle$ and $\langle R_\perp\rangle$, the latter being zero in this phase.
A simple argument that the structure is indeed maximally compact is as follows. It is well known that the 
most compact shape in the two-dimensional continuous space is the circle. For $n$ monomers 
residing in it, $n\approx \pi r^2$, where $r$ is the (dimensionless) radius of this circle.
The usual squared gyration radius is
\begin{equation}
\label{eqcirc}
{R_{\rm gyr}^{\rm circ}}^2 (\approx R_\parallel^2) = 
\frac{1}{\pi r^2}\int\limits_{r'\le r} d^2r' {r'}^2 = \frac{1}{2}r^2
\end{equation}
and therefore $R_{\rm gyr}^{\rm circ}\approx \sqrt{n/2\pi}\approx 5.34$ for $n=N=179$.
Indeed, this is close to the value $R_\parallel\approx 5.46$ of the ground-state conformation we 
identified in phase AC1.
Note that the most compact shape in the simple lattice polymer model we used in our study is
a square and not a discretized circle.~\cite{rem3} 

Near $T\approx 0.3$, the strong layering transition from AC1 to AGe is accompanied by an 
immediate decrease of 
$\langle R_\parallel\rangle$, while $\langle R_\perp\rangle$ rapidly increases from zero to 
about $0.5$ which is exactly the gyration radius (perpendicular to the layers) of a two-layer system,
where both layers cover approximately the same area. Note that the single layers are still compact, but not
maximally. Applying the same approximation as in Eq.~(\ref{eqcirc}), the planar gyration radius 
for each of the two layers is now (with $n\approx N/2$) 
$R_{\rm gyr}^{\rm circ}\approx 3.77$, while we measured in this phase (AGe) $R_\parallel\approx 4.05$. 
This separates the subphase AGe from the other two-layer pseudo-phase AC2d in Fig.~\ref{figpd}, where
the dominating conformation has perfect two-layer (lattice) structure with $R_\parallel\approx 3.85$
(this is the same 2\% difference between continuous and lattice calculation for perfect shapes as above). 
We will discuss the conformational peculiarities in the following in more detail.
The subphase transition from AGe to AG near $T\approx 0.7$ 
is accompanied by a further decrease of $\langle R_\parallel\rangle$
whereas $\langle R_\perp\rangle$ increases, i.e., the height of the surface-adsorbed globule
increases at the expense of the width. This tendency is stopped when approaching the 
transition ($T\approx 2.0$) from the globular regime AG to the phase of expanded, but still 
adsorbed conformations. While $\langle R_\perp\rangle$ remains widely constant (the fluctuation
does not signalize any transition), the polymer strongly extends in the directions parallel
to the substrate, as indicated by the peak of $d\langle R_\parallel\rangle/dT$. After unbinding
from the substrate, parallel and perpendicular gyration radii behave widely isotropically 
($\langle R_\perp\rangle^2\approx \langle R_\parallel\rangle^2/2\approx \langle R_{\rm gyr}\rangle^2/3$) 
as the influence of the isotropy-disturbing walls is weak in this regime.
\section{The whole picture: The free-energy landscape}
\label{secfree}
\begin{figure}
\centerline{\epsfxsize=8.8cm \epsfbox{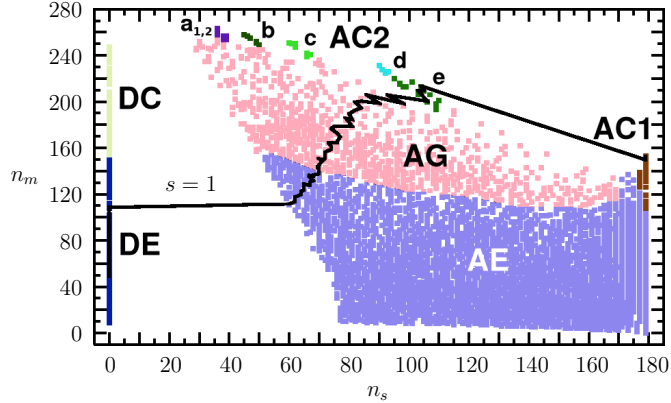}}
\caption{\label{figmap179} (Color online) 
Map of all minima of the contact free energy $F_{T,s}(n_s,n_m)$ in the parameter intervals 
$T\in [0,10]$ and $s\in [-2,10]$ for the 179-mer. The solid line connects the
free-energy minima taken by the polymer in solvent with $s=1$ by increasing the
temperature from $T=0$ to $T=5$ and thus symbolizes its ``path'' through the 
free-energy landscape. The solid line is only a guide to the eyes.} 
\end{figure}

It was shown in Sect.~\ref{ssecfl} that the contact numbers $n_s$ and $n_m$
are unique system parameters for the pseudo-phase identification of the hybrid
system. We define the restricted partition sum for a macrostate with $n_s$ surface
contacts and $n_m$ monomer-monomer contacts by
\begin{eqnarray}
\label{eqrpartsum}
Z_{T,s}(n_s,n_m)&=&\sum\limits_{n'_s,n'_m} \delta_{n'_s n_s}\delta_{n'_m n_m}g_{n'_s n'_m} 
e^{-E_s(n'_s, n'_m)/k_BT}\nonumber \\
&=&g_{n_s n_m}e^{-E_s(n_s, n_m)/k_BT},
\end{eqnarray}
such that $Z=\sum_{n_s,n_m}Z_{T,s}(n_s,n_m)$.
Assuming as usual that the dominating macrostate is given by the minimum of the free energy
as a function of appropriate system parameters, it is useful to define the specific contact free
energy as a function of the contact numbers $n_s$ and $n_m$,
\begin{eqnarray}
\label{eqrfreeb}
F_{T,s}(n_s,n_m)&=& -k_BT\ln g_{n_s n_m}e^{-E_s(n_s, n_m)/k_BT}\\
&=&E_s(n_s,n_m)-TS(n_s,n_m),\nonumber
\end{eqnarray}
identifying $k_B\ln g_{n_s n_m}\equiv S(n_s,n_m)$ as a ``micro-contact'' entropy.
For given external parameters $T$ and $s$, this relation can be used
to determine the minimum of the contact free energy and therefore allows the identification
of the dominant macrostate with respect to the contact numbers. In turn, this quantity allows
for an alternative representation of the pseudo-phase diagram, complementary to the one
shown in Fig.~\ref{figpd} in that it is related to the contact numbers $n_s$ and $n_m$. 
This is done by determining for (in principle) all values of the external parameters
$T$ and $s$ the minima of the contact free energy~(\ref{eqrfreeb}). Then the pair of values 
$n_s$ and $n_m$ of the minimum contact free-energy state are marked in an $n_s$-$n_m$ phase 
diagram. This is shown in Fig.~\ref{figmap179}, where all free-energy minima of the 179-mer
for the parameter set $T\in [0,10]$ and $s\in [-2,10]$ are included and, based on
the arguments of the previous section, differently shaded according to the pseudo-phase
they belong to. The nice thing
of this representation is that it allows the differentiation of continuous and
discontinuous pseudo-phase transitions. 

The first important observation is that the diagram is divided into two separate 
regions, the pseudo-phases of desorbed conformations (DC and DE) and the remaining
different phases of adsorption. The ``space'' in between is blank, i.e., 
none of these (possible) conformations was found to be a free-energy minimum conformation. This
shows that transitions between the adsorbed and desorbed pseudo-phases are always
first-order-like.
It should be noted, that the regime of contact pairs $(n_s,n_m)$ lying {\em above} the 
shown compact phases is forbidden, i.e., conformations with such contact numbers do not 
exist on the sc lattice. 

The second remarkable result is that the pseudo-phases DC, DE, AE, and AG are ``bulky'', while
all AC subphases are highly localized in the plot of the free-energy minima. Comparing with 
Fig.~\ref{figpd}, the conclusion is that 
conformations in the AC phases are energetically favored (more explicitly, for $s/T>0.8$ in 
AC1 and $s/T>2.2$ in the AC2 subphases), while the behavior in the other pseudo-phases is
entropy-dominated: The number of conformations with similar contact numbers in the globular or 
expanded regime is  higher than the rather exceptional conformations in the compact phases,
i.e., for sufficiently small $s/T$ ratios the entropic effect overcompensates the energetic 
contribution to the free energy.

\begin{table}
\caption{\label{tabconf} Representative minimum free-energy examples of conformations in the 
different pseudo-phases of a 179-mer in a cavity. The substrate is shaded in lightgray.}
\begin{tabular}{cccc}\hline\hline
pseudo-phase & example & $n_s$ & $n_m$ \\ \hline
DC & \parbox{3.5cm}{
\centerline{
\epsfxsize = 1.6cm \epsfbox{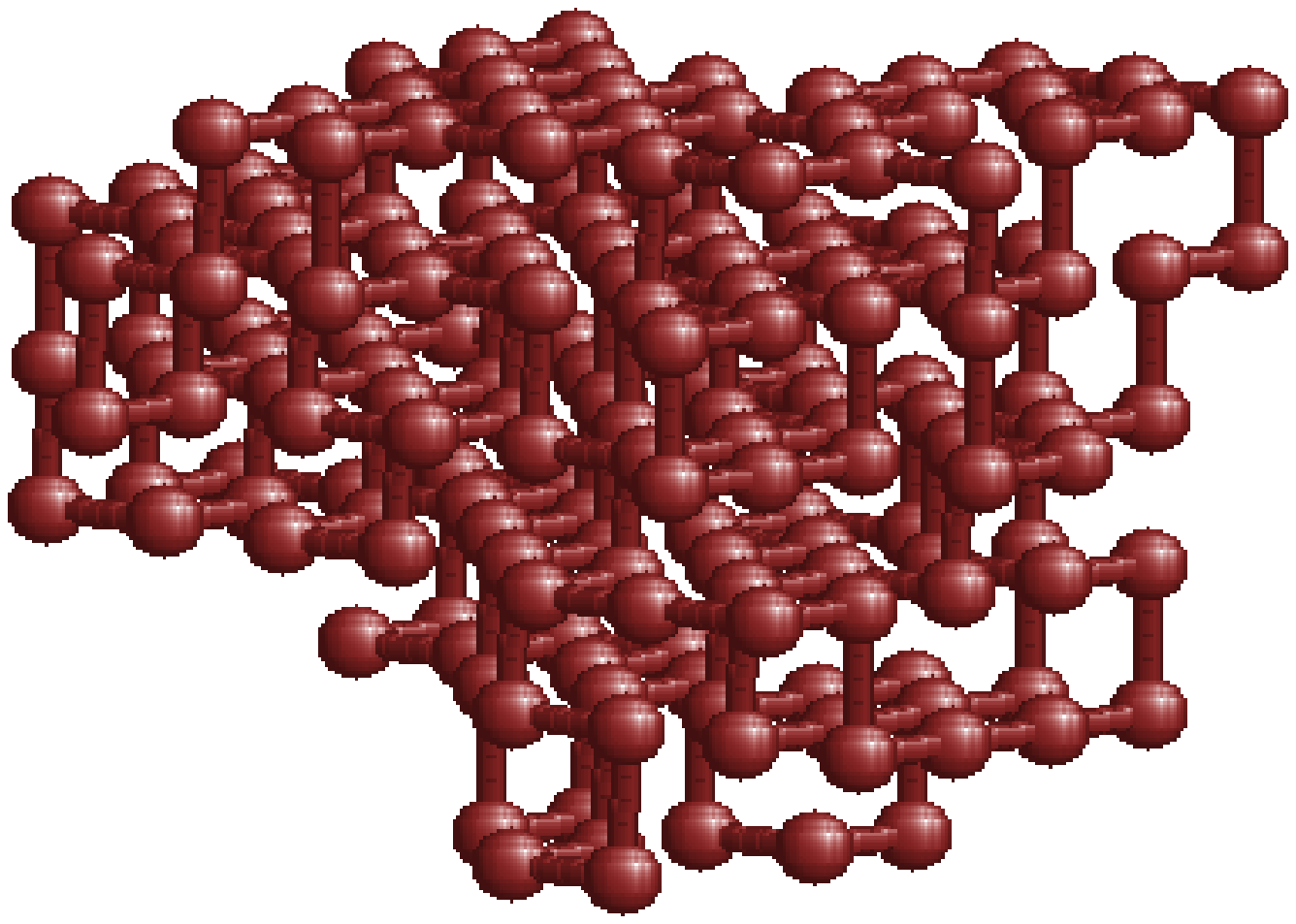}
}} & 0 & 219 \\ \hline
DE & \parbox{3.5cm}{
\centerline{
\epsfxsize = 2.8cm \epsfbox{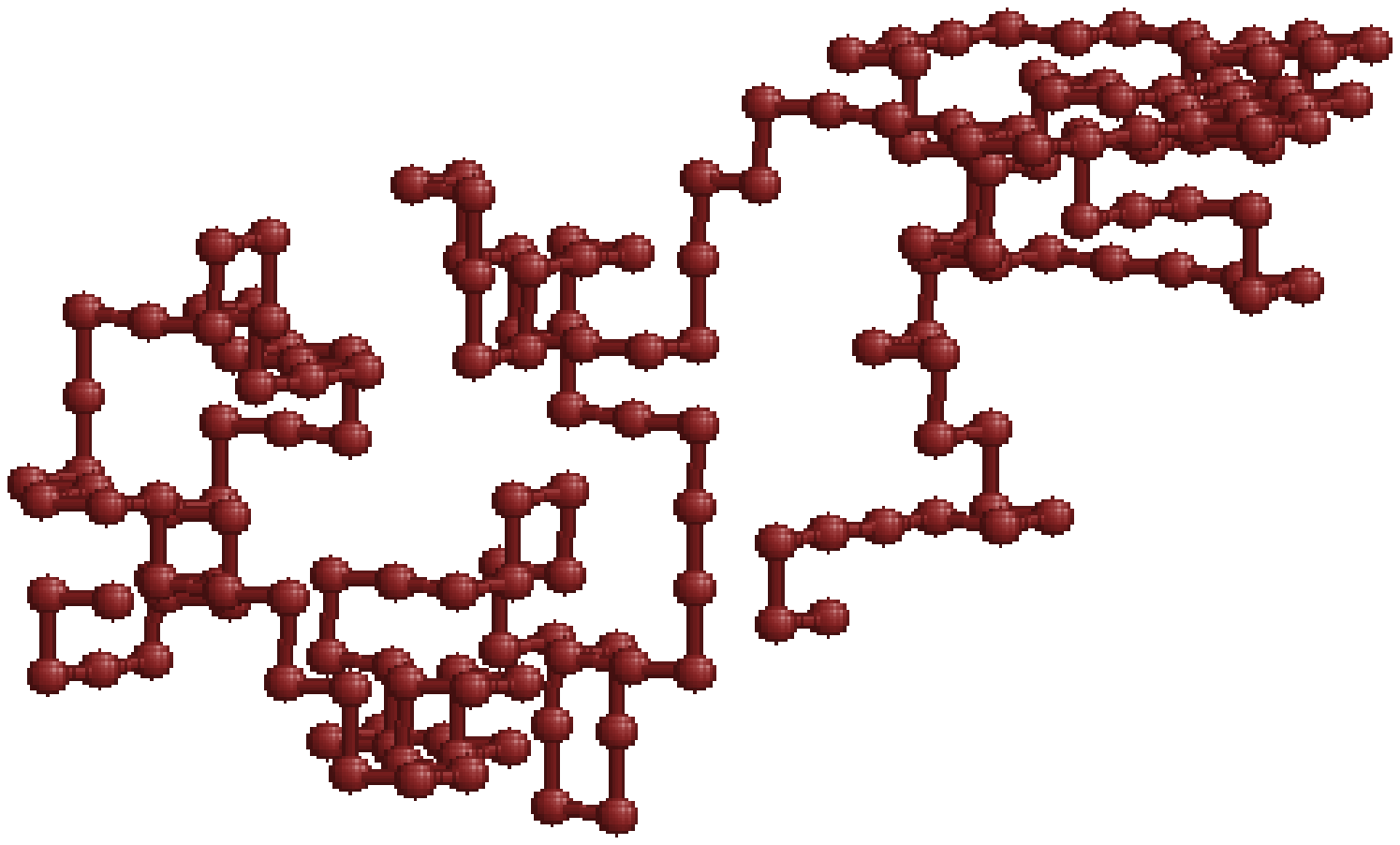}
}} & 0 & 50 \\ \hline
AE & \parbox{4.7cm}{
\centerline{
\epsfxsize = 4.6cm \epsfbox{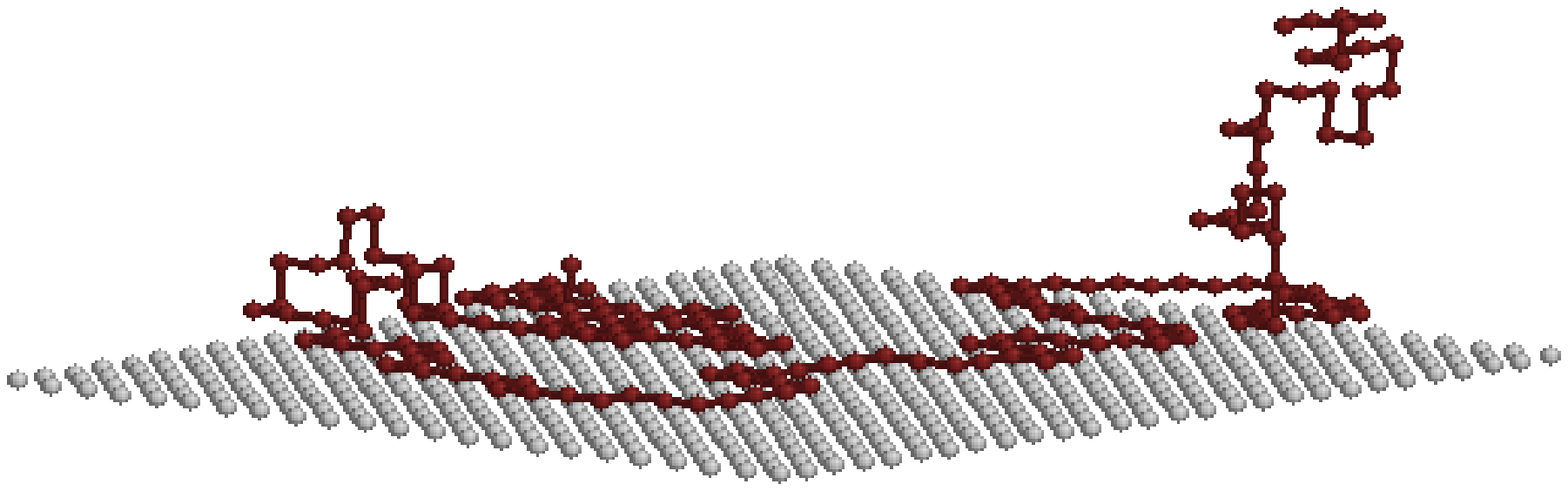}
}} & 135 & 33 \\ \hline
AG & \parbox{3.5cm}{
\centerline{
\epsfxsize = 2.6cm \epsfbox{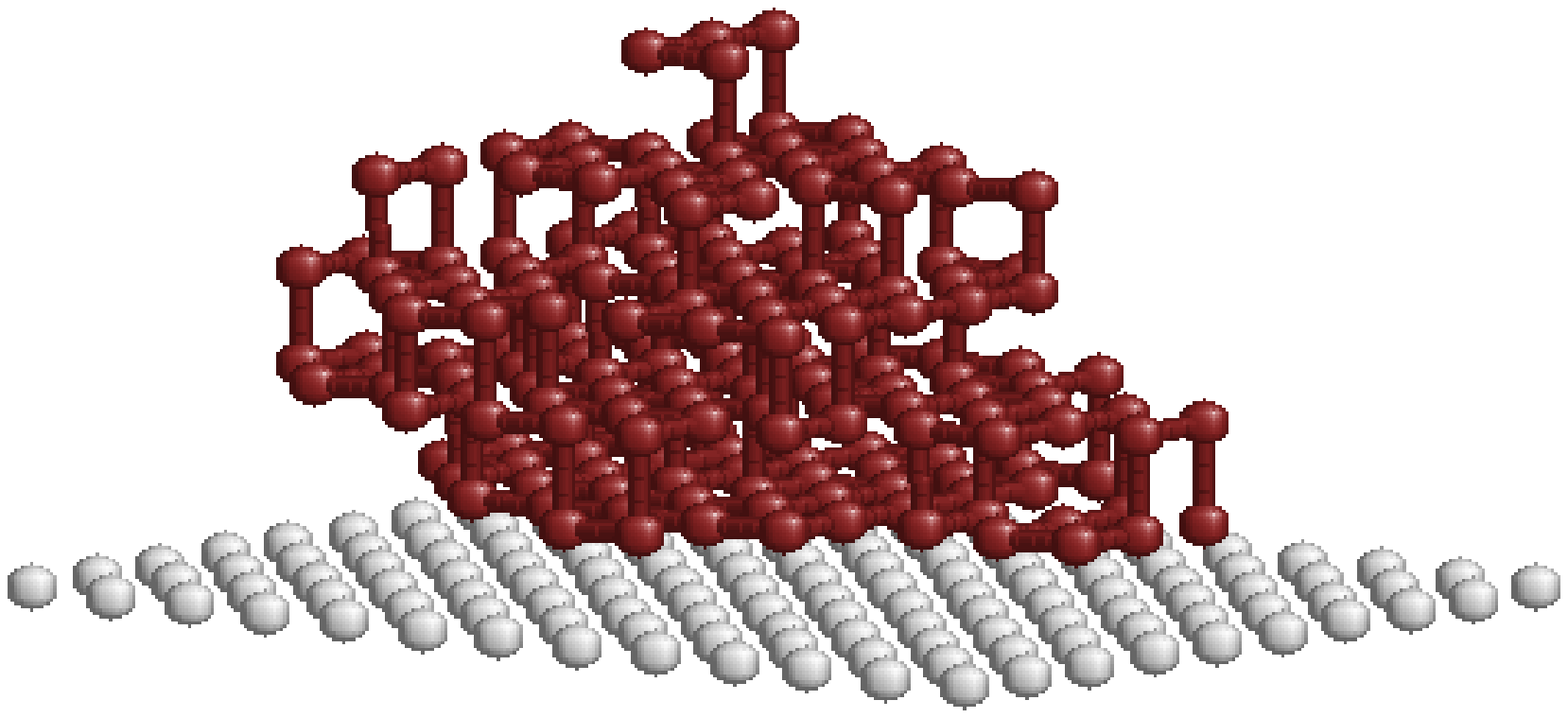}
}} & 49 & 227 \\ \hline
AC2a$_{1}$ & \parbox{3.5cm}{
\centerline{
\epsfxsize = 1.7cm \epsfbox{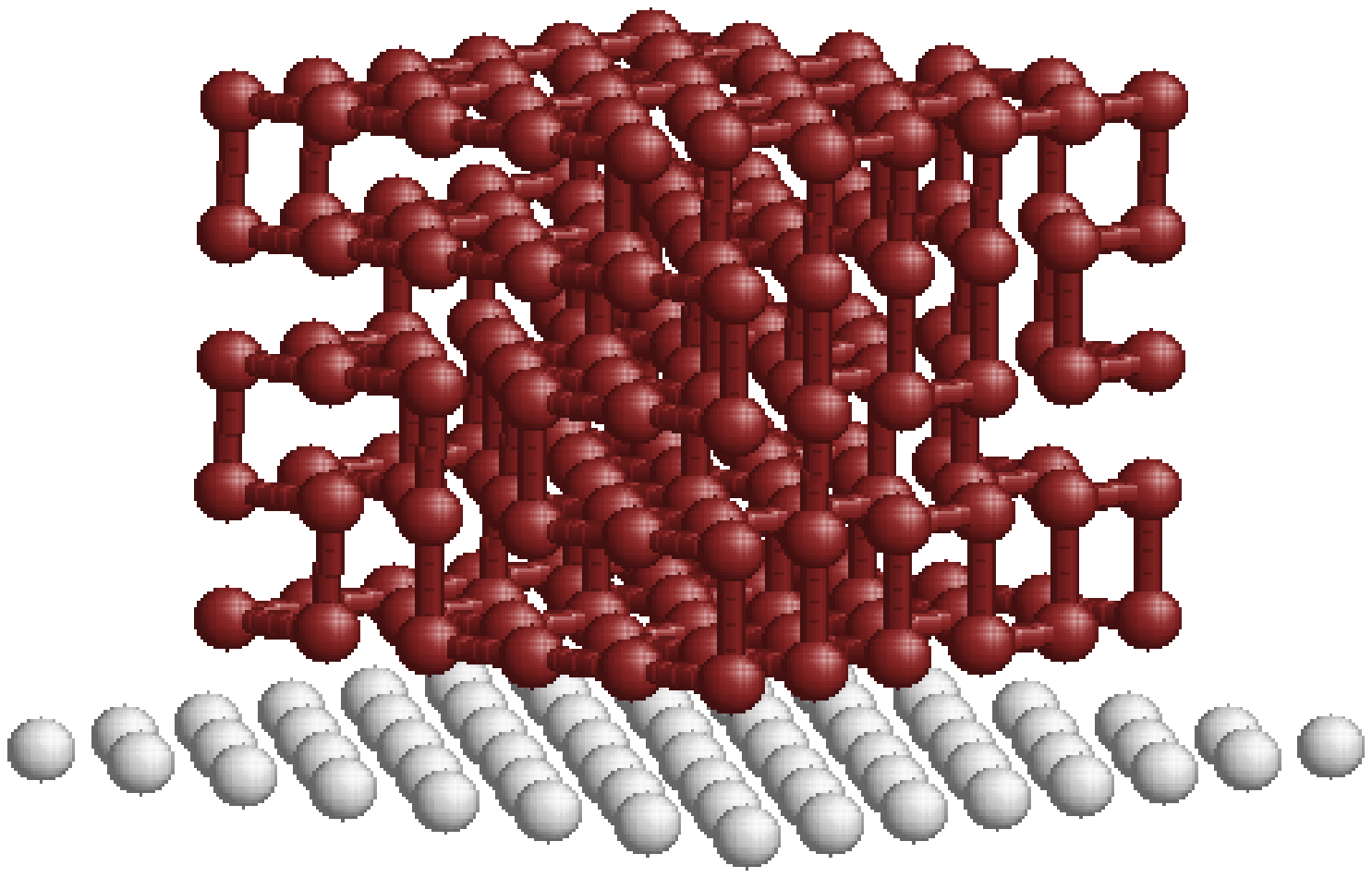}
}} & 36 & 263 \\ \hline
AC2a$_{2}$ & \parbox{3.5cm}{
\centerline{
\epsfxsize = 1.7cm \epsfbox{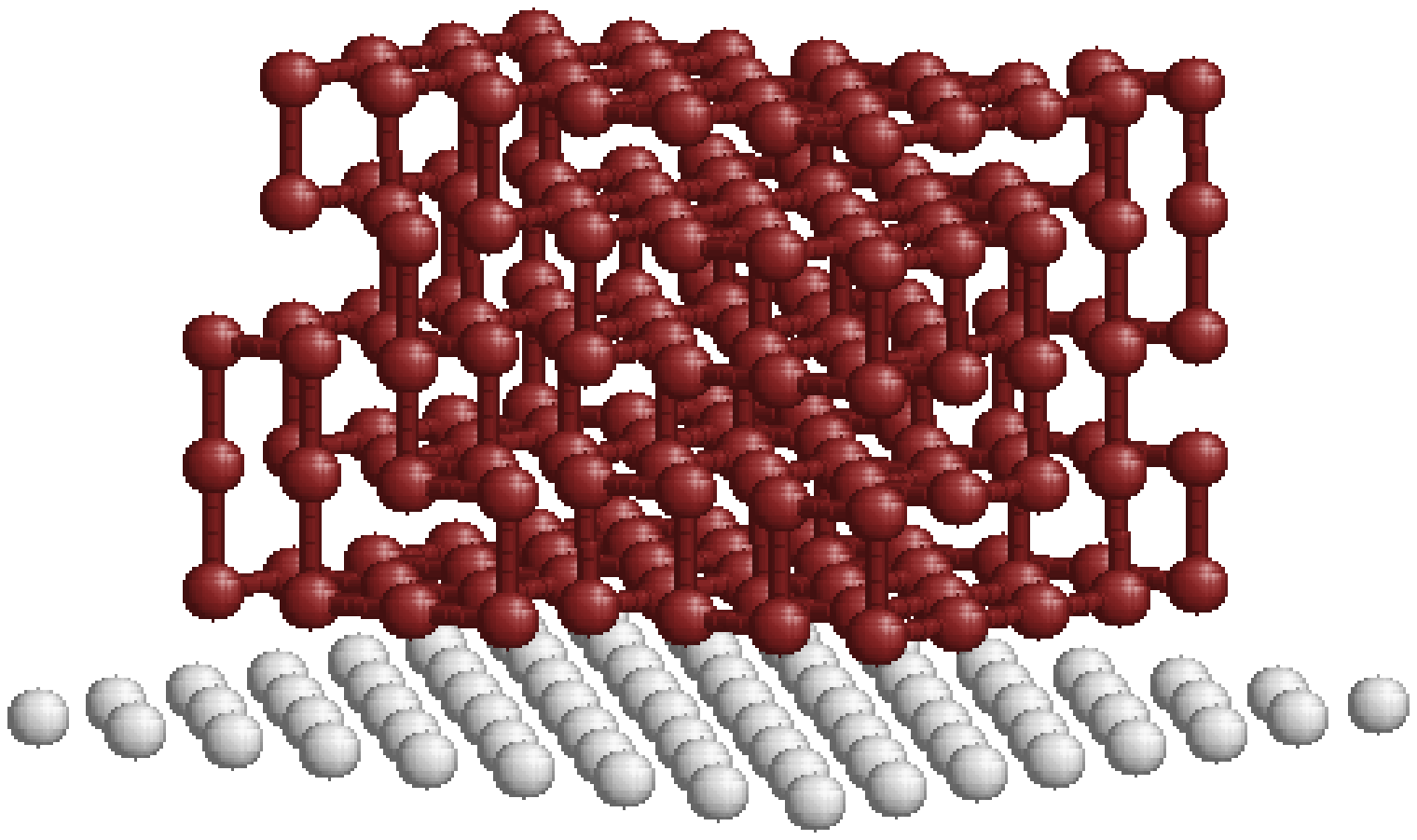}
}} & 39 & 256 \\ \hline
AC2b & \parbox{3.5cm}{
\centerline{
\epsfxsize = 2.2cm \epsfbox{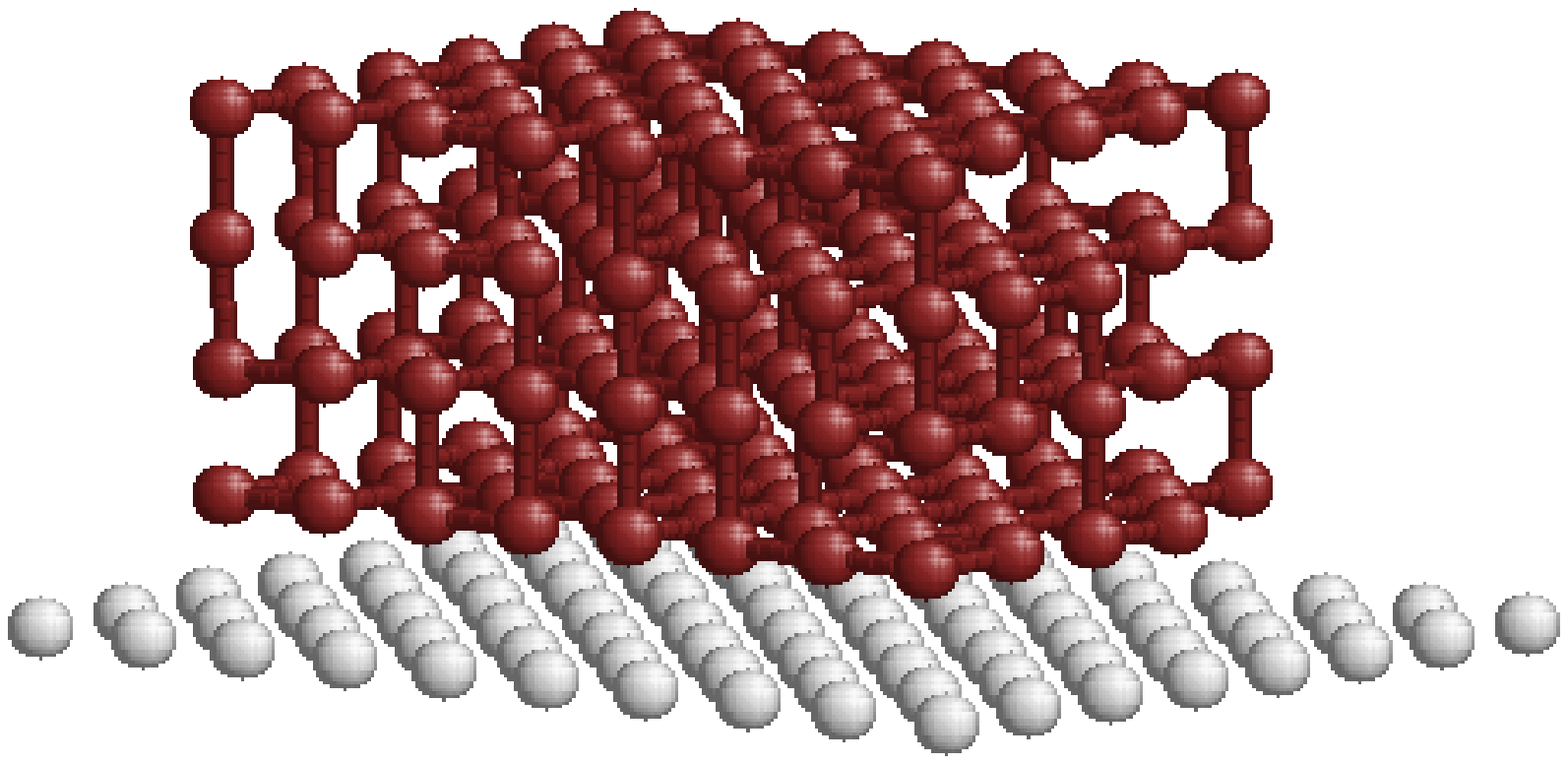}
}} & 46 & 257 \\ \hline
AC2c & \parbox{3.5cm}{
\centerline{
\epsfxsize = 2.2cm \epsfbox{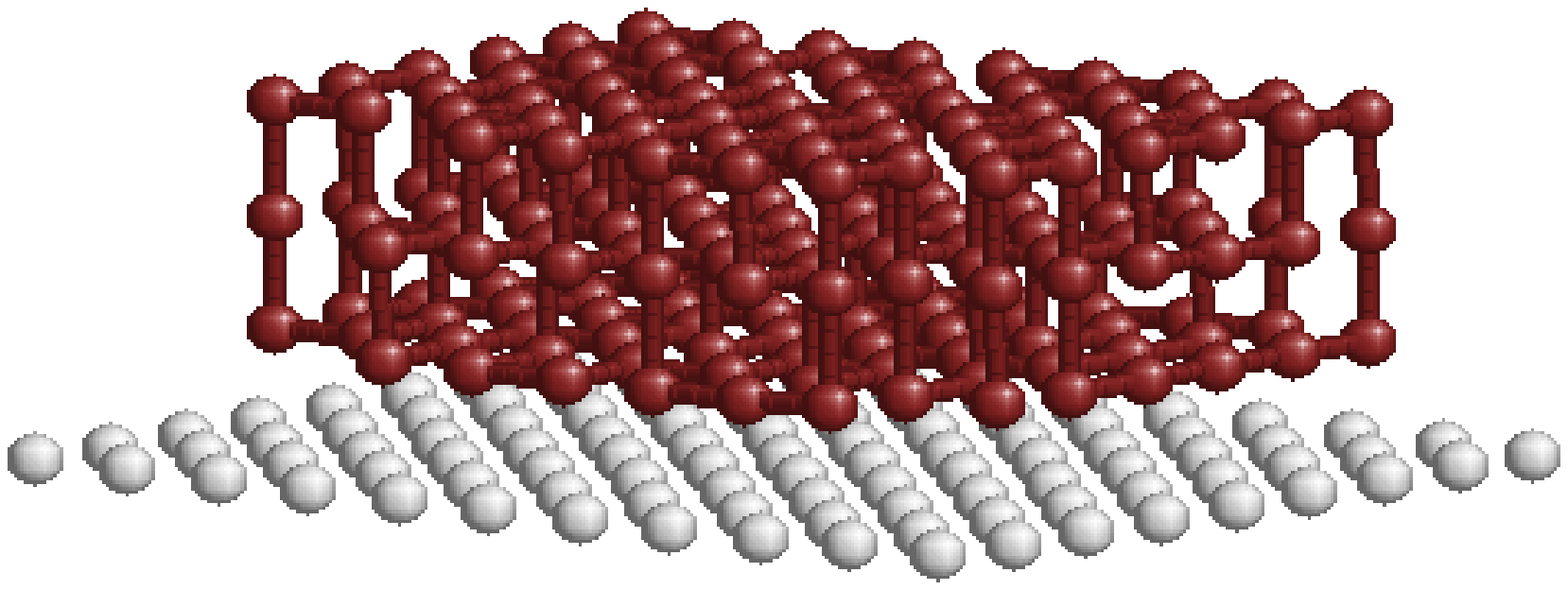}
}} & 60 & 251 \\ \hline
AC2d & \parbox{3.5cm}{
\centerline{
\epsfxsize = 2.6cm \epsfbox{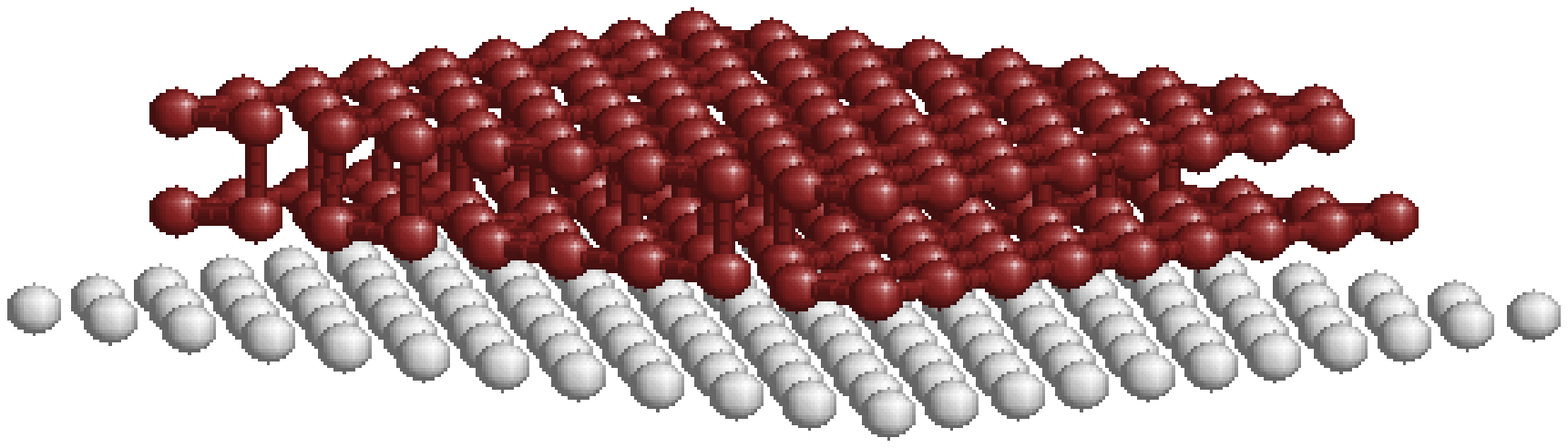}
}} & 90 & 231 \\ \hline
AGe & \parbox{3.5cm}{
\centerline{
\epsfxsize = 3.0cm \epsfbox{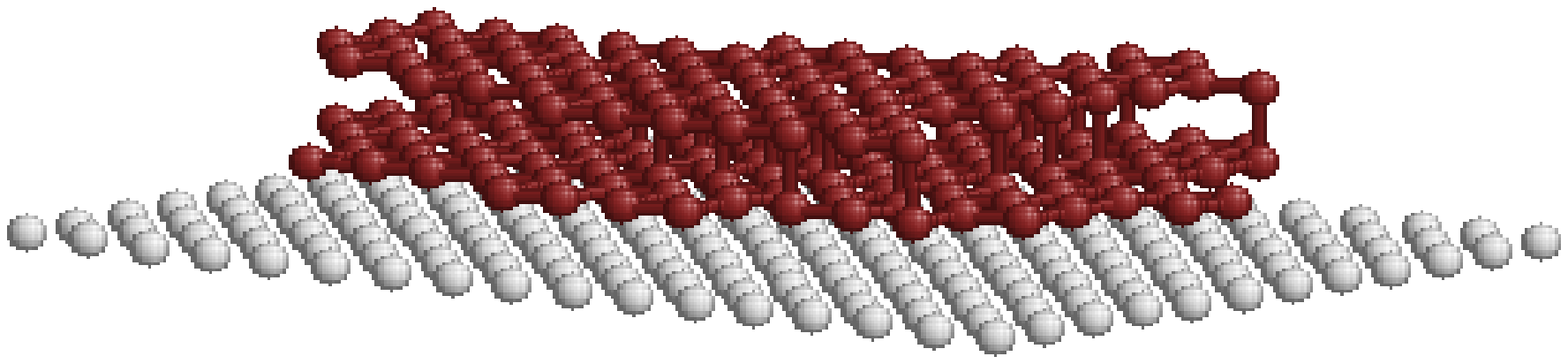}
}} & 103 & 207 \\ \hline
AC1 & \parbox{3.5cm}{
\centerline{
\epsfxsize = 3.4cm \epsfbox{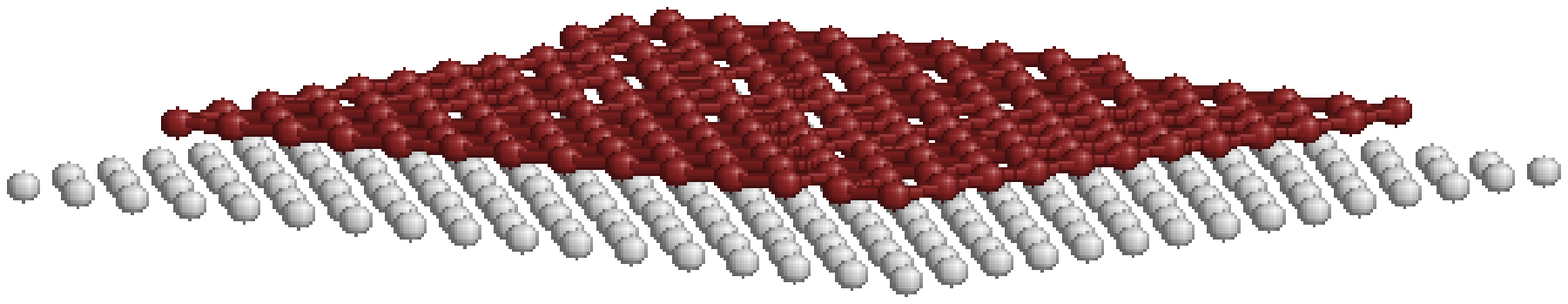}
}} & 179 & 153 \\ \hline\hline
\end{tabular}
\end{table}
The subphases AC2a$_{1,2}$-d are strongly localized, thorn-like ``peninsulas'' standing out from the
AG regime. The discrete number and their separation leads to the conclusion that they have related
structures. Indeed, as can be seen in Table~\ref{tabconf}, where we have listed representative
conformations for all pseudo-phases, the few conformations dominating these subphases exhibit
compact layered structures. The most compact three-dimensional conformation 
with 263 monomer-monomer contacts and 36 surface contacts is favored in subphase AC2a$_1$ and possesses
five layers. Starting from this subphase and increasing the temperature, two things may happen. 
A rather small change is accompanied with the transition to AC2a$_2$, where the number of intrinsic contacts
is reduced but the global five-layer structure remains. On the other hand, passing the transition line
towards AC2b, the monomers prefer to arrange in compact four-layer conformations. 
Advancing towards AC2d, the typical conformations reduce layer by layer in order to increase
the number of surface contacts. In AC2d there are still two layers lying almost perfectly on top
of each other. This is similar in subphase AGe, where also two-layer but less compact conformations 
dominate. In pseudo-phase AC1 only the film-like surface layer remains. The reason for the 
differentiation of the phases AC1 and AC2 of layered conformations is that the transition
from single- to double-layer conformations is expected to be a real phase transition,
while the transitions between the higher-layer AC2 subphases are assumed to disappear in the
thermodynamic limit.~\cite{prellberg1} 

\begin{figure}
\centerline{\epsfxsize=8.8cm \epsfbox{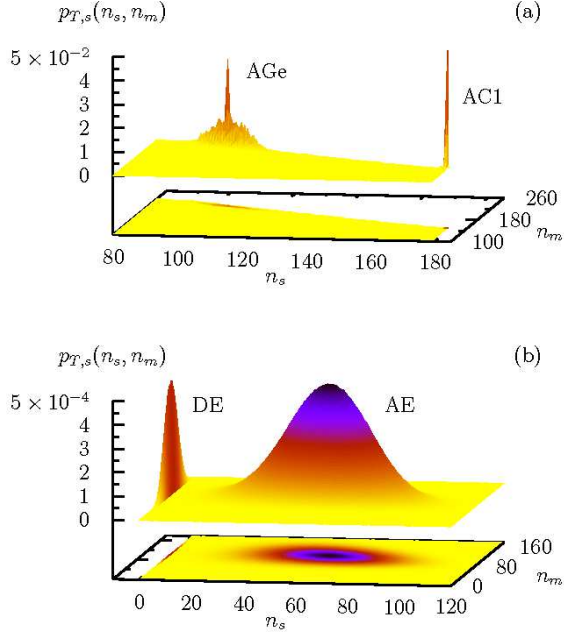}}
\caption{\label{figp179} (Color online)
Probability distributions $p_{T,s}(n_s,n_m)$ for the 179-mer in solvent with $s=1$ 
(a) near the layering transition from AC1 to AGe at $T\approx 0.34$ and
(b) near the adsorption-desorption transition from AE1 to DE at $T\approx 2.44$.
Both transitions are expected to be real phase transitions in the thermodynamic
limit and look first-order-like.}
\end{figure}
As can be seen in Fig.~\ref{figpd}, a transition between AC1 and the phase of adsorbed, 
expanded conformations, AE, is possible. Since these two phases are connected in 
Fig.~\ref{figmap179}, we expect that the transition in between is second-order-like. Indeed,
this transition is strongly related with the {\it two-dimensional} $\Theta$ transition since, 
close to the transition line,
all monomers form a planar (surface-)layer. Similarly, there is also a second-order-like
transition line $s_0(T)$ between AG and AE which separates the regions of poor (AG: $s>s_0$) and  
good (AE: $s<s_0$) solvent. Also, the transition between the desorbed compact (DC) and expanded
(DE) conformations is second-order-like: This transition is strongly related with the well-known
$\Theta$ transition in three dimensions.~\cite{deGennes1} 
Eventually, the transitions from the layer-phases AC2a$_2$, AC2b, AC2c, and AGe to the globular pseudo-phase
AG as well as transitions between pseudo-phases dominated by the same layer type (i.e., between
the two-layer subphases AC2d and AGe, and between the five-layer subphases AC2a$_1$ and AC2a$_2$) 
are expected to be continuous. 

On the other hand, the transitions among the energetically caused compact low-temperature pseudo-phases 
are rather first-order-like, due to their noticeable localization in the map of free-energy minima 
(Fig.~\ref{figmap179}). The possible transitions (see Fig.~\ref{figpd}) are AC2a$_{1,2}$--AC2b,
AC2b--AC2c, and AC2c--AC2d, respectively. Even more interesting, however, are the 
transitions from the single-layer pseudo-phase AC1 to the double-layer subphases AC2d and AGe. 
In the previous sections we already discussed this transition for the special choice $s=1$, where
near $T\approx 0.3$ the fluctuations of the contact numbers and the components of the gyration tensor exhibit
a strong activity. We have included into Fig.~\ref{figmap179} the ``path'' of macrostates the system
passes by increasing the temperature from $T=0$ to $T=5$. At $T=0$ the system is in a film-like,
single-layer state. Near $T\approx 0.3$ it indeed suddenly rearranges into two layers and enters 
subphase AGe in a single step. In Fig.~\ref{figp179}(a) we have plotted the probability distribution
$p_{T,s}(n_s,n_m)$ for $s=1$ and $T= 0.34$ and it can clearly be seen that two distinguished macrostates
coexist.~\cite{j1,j2} Increasing the temperature further, the system undergoes the continuous transitions from AGe via
AG until it unfolds when entering pseudo-phase AE. The system is still in contact with the substrate. 
Close to a temperature $T\approx 2.4$, however, the unbinding of the polymer off the substrate happens
(from AE to DE). Comparing Figs.~\ref{figmap179} and~\ref{figp179}(b), where the probability distribution 
at $T= 2.44$ is shown, we see also a clear indication for a discontinuous transition. Note that we  
consider here the transition state, where the two minima of the free energy coincide~\cite{rem4} (see also the black
dashed line in Fig.~\ref{figpd}) and not the point, where the width of the distribution, i.e., the specific
heat, is maximal. Since the system is finite, the transition temperature ($T\approx 2.8$), as signalled by the 
fluctuations studied in the previous sections, deviates slightly from the transition-state temperature 
reported here.    
\section{Summary}
\label{secsum}
In this paper, we have studied in detail the solubility-temperature (pseudo-)phase diagram of a polymer in a cavity 
with an attractive substrate. We identified the thermodynamic phases of adsorbed compact and expanded (AC, AE) 
and desorbed (DC~\cite{remA}, DE) conformations as well as the previously not yet clearly confirmed phase of adsorbed globules 
(AG). Although the polymer in our study possessed only $N=179$ monomers, these \mbox{(pseudo-)}phases are expected to 
be stable also in the thermodynamic limit $N\to \infty$. 
Other noticeable phase transitions in the compact-globular adsorbed regime (AC1-AC2d, AC1-AGe) are
the energetic layering transitions from film-like surface-layer to double-layer conformations which are also believed
to survive the thermodynamic limit.~\cite{prellberg1} In addition, further subphases of higher-order layers were 
found in low-temperature regions and bad solvent (AC2a$_{1,2}$, AC2b, and AC2c). The most compact three-dimensional conformation found is cube-like
and forms five layers (in subphase AC2a$_1$).

The (pseudo-)phase diagram is based on the specific-heat profile as a function of temperature and 
reciprocal solubility. Although this profile allows for the identification of phases and their boundaries
it does tell little about the conformational transitions between the phases. For this purpose we considered
expectation values and fluctuations for the numbers of monomer-surface contacts, $n_s$, and intrinsic 
monomer-monomer contacts, $n_m$, separately. These contact numbers turned out to be sufficient to describe the 
macrostate of the system and therefore are useful to describe the conformations dominating the different
phases. This view was completed by an exemplified study of the anisotropic behavior of the gyration tensor
components of the polymer parallel and perpendicular to the substrate.

Another central aspect was the classification of the conformational
transitions between the (pseudo-)phases. Based on the contact numbers $n_s$ and $n_m$, we
defined an appropriate free energy and studied the distribution of the minima in the $n_s$-$n_m$ space.
From this kind of free-energy landscape, we found strong indications that the binding-unbinding transitions between
the adsorbed and desorbed phases are first-order-like. This was also observed for the layering transitions.
On the other hand, the transitions across the line separating good and poor solvent, i.e., between
the compact (or globular) and the expanded conformations, are rather second-order-like. This is in coincidence 
with the known behavior of free polymers at the $\Theta$ collapse transition in two and three dimensions. 

Since the experimental equipment and the technological capabilities have nowadays reached an enormous standard
of high single-molecular resolution, we expect that it should be possible to verify experimentally 
not only the existence of the described thermodynamic phases, but also the pseudo-phases being only relevant 
for finite polymers and specific to their lengths. 
\vspace*{10mm}
\section{Acknowledgements}
This work is partially supported by the DFG (German Science Foundation) grant
under contract No.\ JA 483/24-1. Some simulations were performed on the
supercomputer JUMP of the John von Neumann Institute for Computing (NIC), Forschungszentrum
J\"ulich, under grant No.\ hlz11.

\begin{thebibliography}{199}
\bibitem{remA}
In very poor solvent (i.e., very large values of $s$),
the DC phase may degenerate, at least for the finite-length polymers, into separate 
globular and crystalline subphases (see, e.g., Ref~\cite{rampf1}), but this effect is not 
in the focus of the present study of adsorption properties.
\bibitem{rem1}
In our recent study of a 100-mer in a cavity,~\cite{bj1} we have still denoted
AE as AE1 and AG as AE2, because the differentiation between these two phases
was still relatively diffuse due to the small size of the system. For the 
longer chain we are considering here, the identification of these phases is much  
clearer.
\bibitem{rem3}
This is also true for most compact three-dimensional structures which are cube-like.
Therefore, globules in the lattice model are rather cubes than spheres!
\bibitem{rem4}
Strictly speaking, the proper definition of the coexistence point is the temperature,
where the weights under the two peaks are equal, see, e.g., Refs.~[\onlinecite{j1,j2}].
\end{thebibliography}
\end{document}